\documentclass{amsart}
\usepackage{amssymb}
\usepackage{amscd}
\usepackage[arrow,matrix,curve]{xy}
\usepackage{graphicx}
\newcommand\datver[1]{\def\datverp%
   {\par\boxed{\boxed{\text{#1; Run: \today}}}}}

\newcommand\boxb[1]{\square_b}

\newcommand\paperbody%
          {}

\newtheorem{theorem}{Theorem}[subsection]

\newtheorem*{setup}{Basic Setup}
\newtheorem{lemma}[theorem]{Lemma}
\newtheorem{proposition}[theorem]{Proposition}
\newtheorem{corollary}[theorem]{Corollary}

\theoremstyle{remark}
\newtheorem{definition}[theorem]{Definition}
\newtheorem{remark}[theorem]{Remark}

\newcommand\co{\colon\,}
\newcommand\Ind{\operatorname{Ind}}

\newcommand\vol{\text{\textup{vol}}}

\newcommand\Ch{\operatorname{Ch}}

\newcommand\lp{\textup{(}}
\newcommand\rp{\textup{)}}

\newcommand\Fg{\mathfrak{g}}
\newcommand\Fk{\mathfrak{k}}

\newcommand\cH{\mathcal{H}}
\newcommand\cK{\mathcal{K}}
\newcommand\cM{\mathcal{M}}

\newcommand\CC{\mathbb C}

\newcommand\RR{\mathbb R}

\newcommand\ZZ{\mathbb Z}

\newcommand\bbC{\mathbb C}

\newcommand\bbH{\mathbb H}

\newcommand\bbR{\mathbb R}

\newcommand\bbT{\mathbb T}
\newcommand\bbZ{\mathbb Z}
\newcommand\bZ{\bbZ}

\newcommand\cS{\mathcal S}

\newcommand{\TT}{{\mathbb T}}
\newcommand\cont{\text{\textup{cont}}}
\newcommand\Lie{\text{\textup{Lie}}}

\newcommand\Ad{\operatorname{Ad}}

\newcommand\Hom{\operatorname{Hom}}
\newcommand\Aut{\operatorname{Aut}}

\newcommand\PU{\operatorname{PU}}

\newcommand{\im}{\operatorname{im}}
\newcommand{\Br}{\operatorname{Br}}

\newcommand\Ca{$C^*$-algebra}
\newcommand\CTalg{continuous-trace algebra}

\datver{1.4; Revised: 23-1-2004}
\begin{document}
\title[T-duality via noncommutative topology]{T-duality for torus 
bundles\\with H-fluxes\\
via noncommutative topology}

\author{Varghese Mathai}
\address{Department of Pure Mathematics, University of Adelaide,
Adelaide 5005, Australia}
\email{vmathai@maths.adelaide.edu.au}
\author{Jonathan Rosenberg}
\thanks{VM was supported by the Australian Research
Council.
JR was partially supported by NSF Grant DMS-0103647, and thanks
the Department of Pure Mathematics of the University of Adelaide for its
hospitality in January 2004, which made this collaboration possible.}
\address{Department of Mathematics,
University of Maryland,
College Park, MD 20742, USA}
\email{jmr@math.umd.edu}
\dedicatory{\datverp}
\begin{abstract}
It is known that the T-dual of a circle bundle with H-flux 
(given by a Neveu-Schwarz 3-form) is the T-dual
circle bundle with dual H-flux. However, it is also known that
torus bundles with H-flux do not necessarily have a T-dual which is a 
torus bundle. A big puzzle has been to explain these
mysterious
``missing T-duals.''  Here we show that this problem is
resolved using noncommutative topology.  It turns out that
every principal $T^2$-bundle with H-flux does
indeed have a T-dual, but in the missing cases (which we characterize),
the T-dual is non-classical and is
a bundle of noncommutative tori. The duality comes with an
isomorphism of twisted $K$-theories, just as in the classical
case. The isomorphism of twisted cohomology which one gets in
the classical case is replaced by an isomorphism of twisted 
cyclic homology.
\end{abstract}
\maketitle


\section{Introduction}
\label{sec:intro}

An important symmetry of string theories is T-duality, which exchanges
wrapping of fields over a torus with wrapping over the dual torus
\cite{Bus1}, \cite{Bus2}, \cite{AABL}, \cite{Alv12}.
(The exact mathematical meaning of ``dual torus'' is that if $\Lambda$
is a lattice in $\bbR^n$ and $\Lambda^*$ is the dual lattice in the
dual vector space $(\bbR^n)^*$, then $(\bbR^n)^*/\Lambda^*$ is the dual 
torus to $\bbR^n/\Lambda$.) 
Many authors have tried to understand this duality 
from various points of view.
Since Ramond-Ramond (RR) charges are expected to be
represented by classes in $K$-theory (see, e.g., 
\cite{Wa}, \cite{Wb}, \cite{Mkthy}, \cite{MMSi}),
T-duality should come with an isomorphism of $K$-theories
(usually with a degree shift) between a theory and its dual.
The type of $K$-theory appropriate for the situation
(e.g., $K$, $KO$, or $KSp$) depends on the type of
string theory being considered; here we deal with the
type II situation, which leads to complex $K$-theory.
(For a few comments on type I theories, see
section \ref{sec:concl}.)

As pointed out in many contexts (e.g., \cite{SYZ}, \cite{DP}), T-duality
can apply not only to theories over spaces of the form $X\times T^n$,
but also to non-trivial torus bundles, and even to spaces which are
only ``approximately'' of this form, for example, spaces admitting a 
torus action which is generically free. (However, in this paper
we only consider the case of free torus actions.) In addition, it
should apply as well to situations with a non-trivial 
Neveu-Schwarz (NS) 3-form $H$.
In these situations, the H-flux gives rise to a twisting of
$K$-theory, so that one expects an isomorphism of \emph{twisted}
$K$-theories. In its general form, T-duality often involves
a change of topology (see, e.g., \cite{BEMa}, \cite{BEMb}, and 
\cite{BHMb}).

Our initial interest was in trying to explain the
T-duality of torus bundles, in the presence of twisting by
an H-flux, from the perspective of noncommutative topology.
An unexpected byproduct, which we will discuss in section
\ref{sec:ex}, is that we have found that several known
cases of torus bundles with ``missing'' T-duals are in fact
naturally T-dual to \emph{noncommutative} torus bundles, in a sense
we will make precise below. This suggests an unexpected
link between classical string theories and the ``noncommutative'' ones,
obtained by ``compactifying'' along noncommutative tori,
as in \cite{CDS} (cf.\ also \cite[\S\S6--7]{SW}).

Just as a complete characterization of T-duality on circle 
bundles with H-flux is given in  \cite{BEMa} and \cite{BEMb},  
in this paper, we give a complete characterization 
of T-duality on principal $\TT^2$-bundles with H-flux,
Theorem \ref{thm:main}. 
We also describe partial results for T-duality on general 
principal torus bundles with H-flux. The main mathematical
result is a detailed analysis of the equivariant Brauer group
for principal $\TT^2$-bundles, Theorem \ref{thm:n2facts},
which refines earlier results in \cite{CKRW} and \cite{PRW}.
This depends on some explicit calculations of Moore's
``Borel cochain'' cohomology groups.

\section{Preliminaries on noncommutative tori} 
\label{sec:tori}

Here the definition of the (2-dimensional)
noncommutative torus is recalled, cf. 
\cite{Rieffel}. This algebra (stabilized by tensoring
with the compact operators $\cK$) occurs geometrically 
as the foliation algebra associated to Kronecker
foliations on the torus \cite{Connes}. 
It also occurs naturally in the matrix formulation of M-theory as 
the components of Yang-Mills connections 
in the classification of BPS states \cite{CDS}.

For each $\theta \in [0,1]$, the {\em noncommutative torus} 
$A_\theta$ is defined abstractly 
as the $C^*$ algebra generated by two unitaries
$U$ and $V$ in an infinite dimensional Hilbert space 
satisfying the relation
$UV = \exp(2\pi i \theta) VU$. Elements in $A_\theta$ 
can be represented by infinite power series
\begin{equation}\label{schwartz}
f = \sum_{(m, n)\in \ZZ^2} a_{(n, m)} \, U^m V^n
\end{equation} 
where $a_{(m, n)} \in \bbC$ for all $(m, n)\in \ZZ^2$. There is a 
natural smooth subalgebra
$A^\infty_\theta$ called the {\em smooth noncommutative torus}, 
which is defined as those 
elements in $A_\theta$ that can be represented by infinite
power series \eqref{schwartz} with $(a_{(m, n)})\in 
\cS(\ZZ^2)$, the Schwartz space of rapidly decreasing 
sequences on $\ZZ^2$.

$A_\theta$ can also be realized as the crossed product 
$C(\TT) \rtimes_\theta \ZZ$, where the generator of
$\ZZ$ acts on $\TT$ by rotation by the angle $2\pi\theta$.
When $\theta$ is rational, $A_\theta$ is type I, and is
even Morita equivalent to $C(\TT^2)$. However, when $\theta$
is irrational, $A_\theta$ is a simple non-type I {\Ca}.
Because of the realization of $A_\theta$ as a crossed
product by rotation by $2\pi\theta$, the algebra in
this case is often called an \emph{irrational rotation
algebra}. 

Consider the 2 dimensional torus $\TT^2 = \RR^2/\ZZ^2$. For 
each $\theta \in [0,1]$, the
noncommutative torus $A_\theta$ is Morita equivalent to the foliation 
algebra associated to the foliation on $\TT^2$ defined by the 
differential equation 
$dx = \theta \, dy$ on $\TT^2$.

\section{Mathematical framework}
\label{sec:framework}

We begin by explaining the precise mathematical framework
in which we are working. We assume $X$ (which will be
the spacetime of a string theory) is a (second-countable)
locally compact Hausdorff space.  
In practice it will usually be a compact manifold, though 
we do not need to assume this. However it is convenient
to assume that $X$ is finite-dimensional and
has the homotopy type of a finite CW-complex.
(This assumption can be weakened but some finiteness
assumption is necessary to avoid some pathologies. This is
not a problem as far as the physics is concerned.)
We assume $X$ comes with a free
action of a torus $T$; thus (by the Gleason slice theorem
\cite{Gl} the quotient map $p\co X\to Z$
is a principal $T$-bundle.

A \emph{{\CTalg} $A$ over $X$} is a particular type of type I
{\Ca} with spectrum $X$ and good local
structure (the ``Fell condition'' \cite{F}).\footnote{Except
in section \ref{sec:concl} below, all {\Ca}s and Hilbert spaces
in this paper will be over $\bbC$.} 
We will always assume $A$ is
separable; then a basic structure theorem of Dixmier and Douady
\cite{DD} says that after stabilization (i.e., tensoring
by $\cK$, the algebra of compact operators on an infinite-dimensional
separable Hilbert space $\cH$), $A$ becomes \emph{locally}
isomorphic to $C_0(X,\cK)$, the continuous $\cK$-valued
functions on $X$ vanishing at infinity. However, $A$ need not
be \emph{globally} isomorphic to $C_0(X,\cK)$, even after
stabilization. The reason is that a stable {\CTalg} is the
algebra of sections (vanishing at infinity) of a bundle
of algebras over $X$, with fibers all isomorphic to $\cK$.
The structure group of the bundle is $\Aut \cK\cong PU(\cH)$,
the projective unitary group $U(\cH)/\TT$. Since $U(\cH)$ is
contractible and the circle group $\TT$ acts freely on it,
$PU(\cH)$ is an Eilenberg-MacLane $K(\bbZ,2)$-space,
and thus bundles of this type are classified by
homotopy classes of continuous maps from $X$ to
$BPU(\cH)$, which is a $K(\bZ,3)$-space, or
in other words by $H^3(X,\bZ)$. Alternatively,
the bundles are classified by
$H^1(X,\,\underline{PU(\cH)})$, the sheaf cohomology
of the sheaf $\underline{PU(\cH)}$ of germs of continuous
${PU}$-valued functions on $X$, where the transition
functions of the bundle naturally live.  But
because of the exact sequences in sheaf cohomology
\[
0=H^1(X,\,\underline{U(\cH)})\to
H^1(X,\,\underline{PU(\cH)}) \to H^2(X,\,\underline{\TT}) \to 0
\]
and
\[
0=H^2(X,\,\underline{\bbR})\to
H^2(X,\,\underline{\TT})  \to H^3(X,\bbZ) \to H^3(X,\,\underline{\bbR})=0,
\]
the bundles are classified
by $H^2(X,\underline{\TT})\cong H^3(X,\bbZ)$ \cite[\S1]{RosCT}.
Hence stable isomorphism classes of {\CTalg}s over $X$
are classified by the \emph{Dixmier-Douady class} in $H^3(X,\bbZ)$.
It turns out that {\CTalg}s over $X$, modulo
Morita equivalence over $X$, naturally form a group under the
operation of tensor product over $C_0(X)$, called the
\emph{Brauer group} $\Br(X)$, and that this group is
isomorphic to $H^3(X,\bbZ)$ via the Dixmier-Douady class.

Given an element $\delta\in H^3(X,\bbZ)$, we denote by
$CT(X,\delta)$ the associated stable {\CTalg}. (Thus if 
$\delta=0$, this is simply $C_0(X,\cK)$.) The 
(complex topological) $K$-theory
$K_\bullet(CT(X,\delta))$ is called the \emph{twisted
$K$-theory} \cite[\S2]{RosCT} of $X$ with twist $\delta$, denoted
$K^{-\bullet}(X,\delta)$. When $\delta$ is torsion,
twisted $K$-theory had earlier been considered by
Karoubi and Donovan \cite{KD}. When $\delta=0$, twisted $K$-theory 
reduces to ordinary $K$-theory (with compact supports).

Now recall we are assuming $X$ is equipped with a free $T$-action
with quotient $X/T=Z$. (This means our theory is ``compactified
along tori'' in a way reflecting a global symmetry
group of $X$.) In general, a group action on $X$ need not
lift to an action on $CT(X,\delta)$ for any value
of $\delta$ other than $0$, and even when such a lift
exists, it is not necessarily essentially unique.
So one wants a way of keeping track of what lifts are
possible and how unique they are.
The correct generalization of $\Br(X)$ to the equivariant
setting is the \emph{equivariant Brauer group} defined in
\cite{CKRW}, consisting of equivariant Morita
equivalence classes of {\CTalg}s over $X$ equipped with
group actions lifting the action on $X$. By \cite[Lemma 
3.1]{CKRW},
two group actions on the same stable {\CTalg} over $X$
define the same element in the equivariant Brauer group
if and only if they are outer conjugate. (This implies
in particular that the crossed products are isomorphic.)
Now let $G$ be the universal cover of the torus $T$,
a vector group. Then $G$ also acts on $X$ via the
quotient map $G\twoheadrightarrow T$ (whose
kernel $N$ can be identified with the free abelian
group $\pi_1(T)$). In our situation there are three Brauer
groups to consider: $\Br(X)\cong H^3(X,\bbZ)$,
$\Br_T(X)$, and $\Br_G(X)$.  It turns out, however, that
$\Br_T(X)$ is rather uninteresting, as it is naturally
isomorphic to $\Br(Z)$ \cite[\S6.2]{CKRW}. 
Again by \cite[\S6.2]{CKRW}, the natural ``forgetful map''
(forgetting the $T$-action) $\Br_T(X)\to \Br(X)$ 
can simply be identified with
$p^*\co \Br(Z)\cong H^3(Z,\bbZ) \to H^3(X,\bbZ)\cong \Br(X)$.

Finally, we can summarize what we are interested in.
\begin{setup}
\label{assumptions}
A spacetime $X$ compactified over a torus $T$ will
correspond to a space $X$ {\lp}locally compact, finite-dimensional
homotopically finite{\rp} equipped with a free
$T$-action. The quotient map $p\co X\to Z$
is a principal $T$-bundle.  The NS 3-form
$H$ on $X$ has an integral cohomology class $\delta$ which
corresponds to an element of $\Br(X)\cong H^3(X,\,
\bbZ)$. A pair $(X,\delta)$ will be a candidate for
having a $T$-dual when the $T$-symmetry of $X$ lifts
to an action of the vector group $G$ on $CT(X,\delta)$, or in
other words, when $\delta$ lies in the image of the
forgetful map $F\co \Br_G(X)\to \Br(X)$.
\end{setup}

\section{Structure of the equivariant Brauer group
and T-duality}
\label{sec:eqBr}

Throughout this section, the above Basic Setup will be in force.
We let $n=\dim T$, the dimension of the tori involved.

\subsection{Review of the case $n=1$}
\label{sec:n1}

The case $n=1$ was treated in \cite[Theorem 4.12]{RR}, from a purely
{\Ca}ic perspective, in \cite{BEMa}, from a combined
mathematical and physical perspective, and in \cite{BEMb}
from a more physical point of view. In this case, $G=\bbR$,
$T=\TT=\bbR/\bbZ$, and $N=\bbZ$.  By \cite[Corollary 6.1]{CKRW},
the forgetful map $F\co \Br_G(X)\to \Br(X)$ is an isomorphism,
and thus \emph{every} $\delta\in H^3(X,\,\bbZ)$ is dualizable,
in fact in a unique way. It is proven in \cite{BEMa} that the
T-dual of the pair $(p\co X\to Z,\,\delta)$ is a pair
$(p^\#\co X^\#\to Z,\,\delta^\#)$, where $X^\#$ is
another principal circle bundle over $Z$ and $\delta^\#
\in H^3(X^\#,\,\bbZ)$. Furthermore, there is a beautiful
symmetry in this situation.  Principal $\TT$-bundles over
$Z$ are classified by their Euler class in $H^2(Z,\,\bbZ)$,
or equivalently by the first Chern class of the associated 
complex line bundle. So let $[p],\,[p^\#]\in
H^2(Z,\,\bbZ)$ be the
characteristic classes of the two circle bundles. One has
\begin{equation}
\label{eq:n1duality}
p_!(\delta)=[p^\#],\quad (p^\#)_!(\delta^\#)=[p],
\end{equation}
where $p_!$ and $(p^\#)_!$ are the push-forward maps
in the Gysin sequences of the two bundles.  At the level of
forms, $p_!$ and $(p^\#)_!$ are simply ``integration over
the fiber,'' which reduces the degree of a form by one.

Furthermore, the crossed product $CT(X,\delta)\rtimes \bbR$
is isomorphic to $CT(X^\#,\delta^\#)$, and
$CT(X^\#,\delta^\#)\rtimes \bbR$ is isomorphic to 
$CT(X,\delta)$. In fact, the $\bbR$-action on $CT(X^\#$,
$\delta^\#)$ may be chosen to be the dual action on the
crossed product. If one takes the crossed product 
$CT(X,\delta)\rtimes \bbZ$ by the $\bbR$-action restricted to
$\bbZ=\ker (\bbR\to\TT)$, or the similar
crossed product $CT(X^\#,\delta^\#)\rtimes \bbZ$, 
the result is
\[
CT\bigl(X\times_Z X^\#, \,p^*(\delta^\#)=
(p^\#)^*(\delta)\bigr).
\]
Thus one obtains a commutative diagram of principal $\TT$-bundles
\begin{equation}
\label{eq:fibdiag}
\xymatrix{
 & X\times_Z X^\# \ar[ld]_{p^*(p^\#)} \ar[rd]^{(p^\#)^*(p)} & \\
X \ar[rd]_p& & X^\# \ar[ld]^{p^\#}\\
 & Z & .}
\end{equation}

Finally, we get the desired isomorphisms of twisted
$K$-theory and of twisted homology by using the above
results on crossed products and applying
Connes' Thom isomorphism theorem \cite{CThom}
and its analogue in cyclic homology, due to Elliott, Natsume, 
and Nest \cite{ENN}.
The final result, found in \cite{BEMa}, is a commutative
diagram
\begin{equation} \label{eq:TdualKthyn1}
\xymatrix{
K^{\bullet+1}(X,\delta) \ar[r]^{T_!}_\cong \ar[d]^{\Ch}
& K^{\bullet}(X^\#, \delta^\#)  \ar[d]^{\Ch} \\
H^{\bullet +1}(X,\delta) \ar[r]^{T_*}_\cong 
& \, H^{\bullet}(X^\#, \delta^\#) .
}
\end{equation}
Here $\Ch$ is the Chern character, which is an isomorphism
after tensoring with $\bbR$, and homology should be
$\bbZ/2$-graded (i.e., we lump together all the even cohomology
and all the odd cohomology).
Since this duality interchanges even and odd $K$-theory,
it also exchanges type IIa and type IIb string theories.

\subsection{Features of the general case}
\label{sec:nbig}

We return again to the Basic Setup in section
\ref{assumptions}, but now with $T$ a torus
of arbitrary dimension $n$, so $G\cong \bbR^n$. When $n>1$,
it is no longer true that the forgetful map $F:\Br_G(X)\to
\Br(X)$ is an isomorphism. However, some facts about this map
are contained in \cite{CKRW} and in \cite{PRW}. We
briefly summarize a few of these results, specialized to
the case where $G$ is connected (which forces $G$ to act
trivially on the cohomology of $X$). So as to avoid confusion
between cohomology of spaces and of topological groups, we have
denoted by $H^\bullet_M(G,A)$ the cohomology of the topological
group $G$ with coefficients in the topological $G$-module
$A$, as defined in \cite{Mcoh}. This is sometimes
called ``Moore cohomology'' or ``cohomology with Borel cochains.''
\begin{theorem}[{\cite[Theorem 5.1]{CKRW}}]
\label{thm:CKRW}
Suppose $G$ is a connected Lie group and
$X$ is a locally compact $G$-space {\lp}satisfying our
finiteness assumptions{\rp}. Then there is an exact sequence
\[
\xymatrix{
\Br_G(X) \ar[r]^F & \ker (d_2)  \ar[r]^(.3){d_3}
& H^3_M(G, C(X,\TT))/\im(d_2')},
\]
where
\[
d_2\co H^3(X,\bbZ)\to H^2_M(G,  H^2(X,\bbZ))
\]
and
\[
d_2'\co H^1_M(G,  H^2(X,\bbZ)) \to H^3_M(G, C(X,\TT)).
\]
In addition, there is an exact sequence
\[
\xymatrix{
H^2(Z,\bbZ) \ar[r]^(.4){d_2''} &
H^2_M(G, C(X,\TT)) \ar[r]^(.65)\xi & \ker F \ar[r]^(.3)\eta
& H^1_M(G, H^2(X,\bbZ)).
}
\]
\end{theorem}

Fortunately, since in our situation $G$ is a vector group
and is thus contractible, $H^\bullet_M(G,A)$ vanishes
when $A$ is discrete, thanks to:
\begin{theorem}[{\cite[Theorem 4]{Wig}}]
\label{thm:Wig}
If $G$ is a Lie group and $A$ is a discrete $G$-module,
then $H^\bullet_M(G,A)$ is canonically isomorphic
to $H^\bullet(BG,\underline{A})$ {\lp}the sheaf cohomology of the
classifying space $BG$ with coefficients in the locally
constant sheaf defined by $A${\rp}.
\end{theorem}
\begin{corollary}
\label{cor:vanishing}
If $G$ is a vector group and if $A$ is a discrete abelian
group on which $G$ acts trivially, then
$H^\bullet_M(G,A)=0$ for $\bullet>0$
\end{corollary}
\begin{proof}
Since the action of $G$ on $A$ is trivial, the sheaf
$\underline{A}$ is constant and can be replaced by
$A$. Since $BG$ is contractible,
$H^\bullet(BG,A)=0$.
\end{proof}

Substituting Corollary \ref{cor:vanishing} into Theorem
\ref{thm:CKRW}, we obtain (since our finiteness assumption on
$X$ implies $H^2(X,\bbZ)$ is countable and discrete):
\begin{theorem}
\label{thm:CKRWimproved}
Suppose $G\cong \bbR^n$ is a vector group and
$X$ is a locally compact $G$-space {\lp}satisfying our
finiteness assumptions{\rp}. Then there is an exact sequence:
\[
\xymatrix@C-.8pc{
H^2(X,\bbZ)\ar[r]^(.4){d_2''} & H^2_M(G, C(X,\TT)) \ar[r]^(.65)\xi & 
\Br_G(X) \ar[r]^F & H^3(X,\bbZ) \ar[r]^(.4){d_3}
& H^3_M(G, C(X,\TT)).}
\]
\end{theorem}

This still leaves one set of Moore cohomology groups to
calculate, namely 
\[
H_M^\bullet(G, C(X,\TT)),\qquad \bullet=2,\,3.
\] 
For purposes
of doing this calculation, it is convenient to use the exact
sequence of $G$-modules:
\begin{equation}
\label{eq:FunctionsOnS1}
0\to H^0(X,\bbZ) \to C(X,\bbR) \to C(X,\TT) \to H^1(X,\bbZ) \to 0.
\end{equation}
This is just the start of the long exact cohomology sequence
for the exact sequence of sheaves
\[
0\to\bbZ \to \underline{\bbR} \to \underline{\TT} \to 0.
\]

Our finiteness assumption on $X$ implies that
the cohomology groups of $X$ are countable and discrete. So
by Corollary \ref{cor:vanishing} again, $H^0(X,\bbZ)$ and
$H^1(Z,\bbZ)$ are cohomologically trivial (for $H^\bullet_M(G,
\text{---})$), and thus 
\begin{equation}
\label{eq:TisR}
H_M^\bullet(G, C(X,\TT))\cong H_M^\bullet(G, C(X,\bbR)),
\qquad \bullet > 1.
\end{equation}
Finally, for computing the latter we can use another result 
from \cite{Wig}:
\begin{theorem}[{\cite[Theorem 3]{Wig}}]
\label{thm:Wigvec}
If $G$ is a Lie group and $A$ is a $G$-module
which is a topological vector space,
then $H^\bullet_M(G,A)$ agrees with ``continuous cohomology''
$H^\bullet_{\cont}(G,A)$,
the cohomology of the complex of continuous cochains.
\end{theorem}

On the other hand, ``continuous cohomology'' for modules
which are topological vector spaces is well studied, so we
can apply:
\begin{theorem}[{``generalized van Est'' 
\cite[Corollaire III.7.5]{Gu}}]
\label{thm:VE}
If $G$ is a connected Lie group and $A$ is a $G$-module
which is a complete metrizable topological vector space,
then $H^\bullet_{\cont}(G,A)$ agrees with the
relative Lie algebra cohomology
$H^\bullet_{\Lie}(\Fg,\Fk; A_\infty)$, where
$\Fg$ is the Lie algebra of $G$, $\Fk$ is the Lie algebra of 
a maximal compact subgroup $K$, and $A_\infty$ is the set
of smooth vectors in $A$ {\lp}for the action of $G${\rp}.
\end{theorem}
\begin{corollary}
\label{cor:homdim}
If $G$ is a vector group with  Lie algebra $\Fg$,
and if $A$ is a $G$-module
which is a complete metrizable topological vector space,
then $H^\bullet_{\cont}(G,A)\cong H^\bullet_{\Lie}(\Fg, A_\infty)$.
In particular, it vanishes for $\bullet >\dim G$.
\end{corollary}
\begin{proof}
For a vector group, $K$ is trivial.  Lie algebra cohomology
is computed from the complex $\Hom (\bigwedge^\bullet \Fg, 
A_\infty)$, which vanishes for $\bullet >\dim G$.
\end{proof}

\subsection{Calculations for the case $n=2$}
\label{sec:n2}

We now specialize our Basic Setup of section \ref{assumptions}
to the case where $n=2$,
i.e., $p\co X\to Z$ is a principal $\TT^2$-bundle, and
now $G = \bbR^2$. We apply Theorem \ref{thm:CKRWimproved}.
But since $H^3_M(G,C(X,\TT))\cong H^3_M(G,C(X,\bbR))$
(by equation \eqref{eq:TisR}), to which we can apply
Theorem \ref{thm:Wigvec} and Corollary 
\ref{cor:homdim}, we obtain:
\begin{proposition}
\label{prop:vanishingpast2}
If $G=\bbR^2$ and $X$ is a $G$-space as above, then
$H^3_M(G,C(X,\TT))$ vanishes and the forgetful map
$F\co \Br_G(X)\to H^3(X,\bbZ)$ is surjective.
\end{proposition}

Furthermore, we can also explicitly compute 
$H^2_M(G,C(X,\TT))$, because of the following:
\begin{lemma}
\label{lem:Liecohom}
If $G=\bbR^2$ and $X$ is a $G$-space as in the Basic
Setup of section \ref{assumptions}, then
the maps $p^*\co C(Z,\bbR)\to C(X,\bbR)$
and ``averaging along the fibers of $p$''
$\int\co C(X,\bbR)\to  C(Z,\bbR)$ {\lp}defined
by $\int f(z) = \int_T f(g\cdot x) \, dg$, where
$dg$ is Haar measure on the torus $T$ and we choose 
$x\in p^{-1} (z)${\rp} induce isomorphisms
\[
H^2_M(G,C(X,\bbR)) \leftrightarrows
H^2_M(G,C(Z,\bbR)) \cong C(Z,\bbR)
\]
which are inverses to one another.
\end{lemma}
\begin{proof}
We apply Theorem \ref{thm:VE}. Note that the $G$-action
on $C(Z,\bbR)$ is trivial, so every element of $C(Z,\bbR)$
is smooth for the action of $G$.  But since $\dim G=2$,
we have for any real vector space $V$ with trivial 
$G$-action the isomorphisms
\[
H^2_M(G,V)\cong H^2_{\Lie}(\Fg,V) \cong  
H^2_{\Lie}(\Fg, \bbR)\otimes V \cong  V,
\]
since $H^2_{\Lie}(\Fg, \bbR)\cong \bbR$ by Poincar\'e duality
for Lie algebra cohomology.

Clearly $\int\!\circ \,p^*$ is the identity on $C(Z,\bbR)$,
so we need to show $p^*\circ \int$ induces an isomorphism
on $C(X,\bbR)$.  The calculation turns out to be local,
so by a Mayer-Vietoris argument we can reduce to the
case where $p$ is a trivial bundle, i.e., $X=(G/N)\times Z$,
with $N=\bbZ^2$ and $G$ acting only on the first factor.
The smooth vectors in $C(X,\bbR)$ for the action of
$G$ can then be identified with $C(Z,C^\infty(G/N))$.
So we obtain
\[
H^2_M\bigl(G,C(X,\bbR)\bigr) 
\cong H^2_{\Lie}\bigl(\Fg,C(Z,C^\infty(G/N))\bigr)
\cong C\Bigl(Z, H^2_{\Lie}\bigl(\Fg, C^\infty(G/N)\bigr)\Bigr),
\]
with the cohomology moving inside since $G$ acts trivially
on $Z$.  However, by Poincar\'e duality
for Lie algebra cohomology,
\[
H^2_{\Lie}\bigl(\Fg, C^\infty(G/N)\bigr)
\cong
H_0^{\Lie}\bigl(\Fg, C^\infty(G/N)\bigr),
\]
which is the quotient of $C^\infty(G/N)$ by all derivatives
$X\cdot f$, $X\in \Fg$ and $f\in C^\infty(G/N)$.
This quotient is $\bbR$ by the de Rham theorem, since
$f(g)\,d\vol(g)$ is exact on $T$ exactly when $f$ is constant.
And it's easy to check that the isomorphism 
$H^2_M\bigl(G,C(X,\bbR)\bigr) \cong C(Z,\bbR)$ 
is induced by $\int$.
\end{proof}

\begin{theorem}
\label{thm:n2facts}
In the  Basic Setup with $n=2$, there is a commutative 
diagram of exact sequences:
\begin{equation*}
\xymatrix{
&H^0(Z,\bbZ) \ar[d]& 0\ar[d] &\\
H^2(X,\bbZ)\ar[r]^(.4){d_2''} & H^2_M(G, C(X,\bbT)) \ar[r]^(.6)\xi
\ar@{.>}[d]_a
& \ker F\ar[r]^\eta \ar[d] & 0 \\
& C(Z, H^2_M(\bbZ^2,\bbT)) \ar[d]_h & \Br_G(X) \ar[l]_(.35)M \ar[d]& \\
 & H^1(Z,\bbZ) \ar[d] & H^3(X,\bbZ)
\ar@{.>}[l]_{p_!} \ar[d]& \\
& 0&0&
}
\end{equation*}
Here $M\co \Br_G(X) \to  C(Z,  H^2_M(\bbZ^2,\bbT))\cong C(Z, \TT)$
is the Mackey obstruction map defined in \cite{PRW}, and
$h\co C(Z, \TT) \to H^1(X,\bbZ)$ is the map sending a continuous
function $Z\to S^1$ to its homotopy class. The definitions
of the dotted arrows will be given in the course of the proof.
\end{theorem}
\begin{proof}
Most of this is immediate from Theorem \ref{thm:CKRWimproved}
together with Proposition \ref{prop:vanishingpast2}. There
are just a few more things to check. First we define the
dotted arrows in the diagram. The arrow $p_!\co H^3(X,\bbZ)\to
H^1(Z,\bZ)$ is ``integration over the fibers'' of the
bundle $T^2\to X\stackrel{p}{\to} Z$; more specifically, it is
the projection of $H^3(X,\bbZ)$ onto $E^{1,2}_\infty$
in the Serre spectral sequence of $p$. Since $E^{1,2}_\infty
\subseteq E^{1,2}_2 = H^1(Z, H^2(T^2,\bZ))$, we can think
of the image as lying in $H^1(Z,\bZ)$. In fact, 
\[
E^{1,2}_\infty \subseteq E^{1,2}_3 =
\ker d_2\co H^1(Z, H^2(T^2,\bZ)) \to
H^3(Z, H^1(T^2,\bZ))\cong H^3(Z, \bbZ^2),
\]
and this map $d_2$ can be identified with cup product
with $[p]\in H^2(Z,\bbZ^2)$.  

Next we define the downward dotted arrow $a$ using
Lemma \ref{lem:Liecohom}. It is simply the
following composite:
\[
H^2_M(G, C(X,\bbT)) 
\xrightarrow[\cong]{\text{eq. \eqref{eq:TisR}}}
H^2_M(G, C(X,\bbR)) 
\xrightarrow[\cong]{\text{Lemma \ref{lem:Liecohom}}}
C(Z,\bbR) \xrightarrow{\exp} C(Z,\bbT).
\]
Exactness of the middle downward sequence
\[
H^0(Z,\bbZ)\to
H^2_M(G, C(X,\bbT)) \stackrel{a}{\to} C(Z,\TT) 
\stackrel{h}{\to} H^1(Z,\bbZ)
\]
follows immediately from \eqref{eq:FunctionsOnS1}
with $X$ replaced by $Z$.

We still need to check commutativity of the squares.
As far as the upper square is concerned, the key fact
is that the restriction map 
\[
\bbR\cong H^2_M(\bbR^2,\bbT)\to
H^2_M(\bbZ^2,\bbT)\cong \bbT
\]
is surjective and can be identified with the
exponential map (see the Hochschild-Serre spectral
sequence 
\[ 
H^p_M(\bbR^2/\bbZ^2, H^q_M(\bbZ^2,\TT)) \Rightarrow
H^\bullet_M(\bbR^2,\TT)
\]
of \cite{Mcoh1} for a method of calculation).  
To check commutativity 
for the upper square, choose a Borel
cocycle $\omega\in Z^2_M(G, C(X,\bbT))$ representing
a class in $H^2_M(G, C(X,\bbT))$. By Lemma \ref{lem:Liecohom}, 
we may assume $\omega$ takes its values in functions
constant on $T$-orbits, i.e., pulled back from
$C(Z,\bbT)$ via $p^*$. As in \cite[Theorem 5.1(3)]{CKRW},
choose a Borel map $u\to U\cM(C_0(X,\cK))$ satisfying
\[
u_s\tau_s(u_t) = \omega(s,t)u_{s+t},\quad s,\,t\in G.
\]
(Here $\tau$ is the action of $G$ on $X$.)
Then by the prescription in \cite{PRW}, $\xi([\omega])$
is given by $C_0(X,\cK)$ with the $G$-action
$s\mapsto (\Ad u_s)\tau_s$.  We need to compute the
Mackey obstruction for the restriction of the action to
$N=\bbZ^2$. But this is just given by $z\mapsto M(u_z)$,
the Mackey obstruction of the projective unitary representation
of $N$ defined by $u$ over a point $z\in Z$. But as the
cocycle of the representation is just $\omega$ restricted
to $z$ (this makes sense since we took $\omega$  to have
values constant on $G$-orbits), we can use the above
fact about restricting Moore cohomology from $G$ to $N$
to deduce that $M(\xi([\omega]))=a([\omega])$.

Finally we need to check commutativity of the bottom 
square. This amounts to showing that if we have an action
$\alpha$ of $G$ on $CT(X,\delta)$ representing an element of
$\Br_G(X)$, then $h\circ M (\alpha) = p_!(\delta)$.
(In the case where $M(\alpha)$ is trivial, this is basically
in \cite{PRW}.)  First of all, we note that $h\circ M (\alpha)$
can only depend on $\delta$, not on the choice of the
action $\alpha$ on $CT(X,\delta)$. The reason is that
any two different actions differ by an element of
$\ker F$, which by the rest of the diagram is in the image of
$H^2_M(G, C(X,\bbT))\cong C(Z,\bbR)$. By commutativity
of the upper square, this only changes $M(\alpha)$ within its
homotopy class. Since we already know
$\Br_G(X)\to H^3(X,\bbZ)$ is surjective,
it follows that $h\circ M$ induces a homomorphism
from $H^3(X,\bbZ)\to H^1(Z,\bbZ)$. This map is trivial
on $p^*(H^3(Z,\bbZ))$, since this part of $H^3(X,\bbZ)$ is
represented by $G$-actions where $N=\bbZ^2$ acts trivially
\cite[\S6.2]{CKRW}. And of course when $N$ acts trivially,
there is no Mackey obstruction.

Next we show that the map $H^3(X,\bbZ)\to H^1(Z,\bbZ)$
induced by $h\circ M$ vanishes on the $E_\infty^{2,1}$
subquotient of the spectral sequence. This consists
(modulo classes pulled back from $H^3(Z,\bbZ)$) of
classes pulled back from some intermediate space
$Y$, where $X\xrightarrow{p_1} 
Y\xrightarrow{p_2} Z$ is some 
factorization of the $T^2$-bundle $p\co X\to Z$ as
a composite of two principal $S^1$-bundles. But given
such a factorization and a class $\delta_Y\in Y$, there
is an essentially unique action of $\bbR$ on 
$CT(Y,\delta_Y)$ compatible with the $S^1$-action on $Y$
with quotient $Z$, because of the results of section
\ref{sec:n1}.  Pulling back from $Y$ to $X$, we get
an action of $\bbR\times \TT$ on $CT(X,p_1^*\delta_Y)$,
or in other words an action of $G$ factoring through
$\bbR\times \TT$. Such an action necessarily has trivial
Mackey obstruction.

So it follows that the map induced by  $h\circ M$ 
factors through the remaining subquotient of $H^3(Z,\bbZ)$,
i.e., $E_\infty^{1,2}$. That says exactly that the map 
factors through $p_!$. By naturality, it must be a 
multiple of $p_!$, and we just need to compute in the case
of a trivial bundle to verify that the multiple is $1$.
Thus the proof is completed with the following
Proposition \ref{prop:Mobstrfortrivbundle}.
\end{proof}
\begin{proposition}
\label{prop:Mobstrfortrivbundle}
Let $p\co X=Z\times \TT^2\to Z$ be a trivial $\TT^2$-bundle,
let $\beta\in H^1(Z,\bbZ)$, and let $\delta=\beta\times\gamma
\in H^3(X,\bbZ)$, where $\gamma$ is the usual generator
of $H^2(\TT^2,\bbZ)\cong\bbZ$. Then there is an action
$\alpha$ of $G=\bbR^2$ on $CT(X,\delta)$, compatible with the
free $\TT^2$-action on $X$, for which $h\circ M(\alpha)
=\beta$.
\end{proposition}
\begin{proof}
Choose a function $f\co Z\to \TT$ with $h(f)=\beta$.
Let  $\cH = L^2(\TT)$ and for $z\in Z$, consider the the projective 
unitary representation $\rho_{f(z)} \co\ZZ^2 \to \PU(\cH)$ 
defined by sending the first generator of $\ZZ^2$ 
to multiplication by the identity
map $\TT\to\TT\hookrightarrow \CC$,
and the second generator to translation by $f(z) \in \TT$.
Then the Mackey obstruction of $\rho_{f(z)}$ is $f(z) \in \TT \cong 
H^2(\ZZ^2, \TT)$.  We can view $\rho$ as a spectrum-fixing
automorphism of $\ZZ^2$ on $C(Z, \cK(\cH))$, which is given at 
the point $z\in Z$ by $\Ad \rho_{f(z)}$. We now let
$(A,\alpha)$ be the $C^*$-dynamical system obtained by inducing up
$\bigl(C(Z, \cK(\cH)),\rho\bigr)$ from $\ZZ^2$ to $\bbR^2$.
More precisely, 
  $$
  \begin{array}{lcl}
  A &=& \Ind_{\ZZ^2}^{\RR^2}\left(C(Z, \cK(\cH)), \rho\right) 
\\[+7pt]
  &=& \left\{ f\co \RR^2 \to C(Z, \cK(\cH)):
  f(t+g) = \rho(g)( f(t)),\;\; t\in \RR^2, g\in\ZZ^2 \right\}.
  \end{array}
  $$
Since $\rho$ acts trivially on the spectrum $Z$ of the inducing
algebra and $A$ is an algebra of sections of a locally trivial
bundle of {\Ca}s with fibers isomorphic to $\cK$,
$A$ is a {\CTalg} having spectrum $Z\times \TT^2$.  There is
a natural action $\alpha$ of $\bbR^2$ on $A$ by translation, and by 
construction, $M(\alpha)=f$.  We just need to compute
the Dixmier-Douady invariant of $A$. We get it by ``inducing
in stages''. Let $B=\Ind_\bbZ^\bbR C(Z, \cK(\cH))$ be the
result of inducing over the first copy of $\bbR$. 
Since the first generator of $\bbZ^2$ was always acting
by conjugation by multiplication by the identity
map $\TT\to\TT$ on $L^2(\TT)$, one can see
that $B$ is a trivial {\CTalg}, viz., $B\cong C_0(Z\times \TT,
\cK(\cH))$. We still have another action of $\bbZ$ on $B$
coming from the second generator of $\bbZ^2$, and
$A=\Ind_\bbZ^\bbR B$, where we induce over the second 
copy of $\bbR$ to get $A$. The action of
$\bbZ$ acts on $B$ is by means of a map $\sigma\co
Z\times \TT \to PU(\cH)=\Aut \cK(\cH)$, whose value
at $(z,t)$ is the product of multiplication by $t$ with
translation by $f(z)$.  Thus the Dixmier-Douady invariant
of $A$ is then $[\sigma]\times c$, where $[\sigma]\in 
H^2(Z\times \TT,\bbZ)$ is the homotopy class of $\sigma\co
Z\times \TT\to PU(\cH)=K(\bbZ,2)$ and $c$ is the usual generator
of $H^1(S^1,\bbZ)$. But $[\sigma]$ is now $h(f)\times c$,
so the Dixmier-Douady class of $A$ is $\beta\times c\times c
=\beta\times\gamma$.
\end{proof}

\subsection{Applications to T-duality}
\label{sec:mainthm}

Now we are ready to apply Theorem \ref{thm:n2facts} to
T-duality in type II string theory. First we need a
definition.
\begin{definition}
\label{def:classicaldual}
Let $p\co X\to Z$ be a principal $T$-bundle
as in the Basic Setup of section \ref{assumptions},
and let $\delta\in H^3(X,\bbZ)$. We will say that the
pair $(p,\delta)$ has a 
\emph{classical T-dual} if there
is an element $[A,\alpha]$ of $\Br_G(X)$, with $A$
a {\CTalg} over $X$ with Dixmier-Douady class $\delta$,
and with $\alpha$ an action of $G$ on $A$ inducing the
given free action of $T=G/N$ on $X$, such that the crossed product
$A\rtimes G$ is again a {\CTalg} over some other principal
torus bundle over $Z$, with the dual action of
$\widehat G$ inducing the bundle projection to $Z$.
\end{definition}

This definition is essentially equivalent to that in
\cite{BHMb}; we will say more about this later in
Remark \ref{rem:remdualH}.

The following is the main result of this paper.

\begin{theorem}\label{thm:main}
Let $p\co X\to Z$ be a principal $\TT^2$-bundle
as in the Basic Setup of section \textup{\ref{assumptions}}. Let
$\delta \in H^3(X, \bbZ)$ be an ``H-flux'' on $X$. Then:
\begin{enumerate}
\item If $p_! \delta = 0 \in H^1(Z, \ZZ)$, then there 
is a {\lp}uniquely determined{\rp} classical T-dual to
$(p,\delta)$, consisting of
$p^\#\co X^\# \to Z$, which is a another principal $\TT^2$-bundle
over $Z$, and $\delta^\# \in H^3(X^\#, \bbZ)$,
the ``T-dual H-flux'' on  $X^\#$. One obtains
a picture exactly like equation \textup{\eqref{eq:fibdiag}}. 
\item If $p_! \delta \ne 0 \in H^1(Z, \ZZ)$, then 
a classical T-dual as above does \emph{not} exist.
However, there is a ``nonclassical''
T-dual bundle of noncommutative tori over $Z$. It is not
unique, but the non-uniqueness does not affect its
$K$-theory.
\end{enumerate}  
\end{theorem}

\begin{proof}
By Theorem \ref{thm:n2facts}, the map $F\co\Br_G(X)\to H^3(X,\bZ)$
is always surjective.  This will be the key to the proof.

First consider the case when $p_! \delta = 0 \in H^1(Z, \ZZ)$.
This case is considered in \cite{BHMb}, but we will redo the
results using Theorem \ref{thm:n2facts}.  By commutativity
of the lower square, we can lift $\delta\in H^3(X,\bbZ)$ to
an element $[CT(X,\delta), \alpha]$ of $\Br_G(X)$ with
$M(\alpha)$ homotopically trivial.  Then by using
commutativity of the upper square in Theorem \ref{thm:n2facts},
we can perturb $\alpha$, without changing $\delta$, so
that $M(\alpha)$ actually vanishes. Once this is
done, the element we get in $\Br_G(X)$ is actually unique.
On the one hand, this can be seen from \cite[Lemma 1.3]{PRW} and
\cite[Corollary 5.18]{PRW}. Alternatively, it can be
read off from Theorem \ref{thm:n2facts}, since any
two classes in $\ker M$ mapping to the same $\delta\in 
H^3(X,\bbZ)$ differ by the image under $\xi$ of something
in $\ker a$. Thus they differ by the image under $\xi$ of 
an $\bbZ$-valued cocycle, which is trivial since
such a cocycle exponentiates to the trivial cocycle
with values in $\bbT$, and this is all that is used
in the construction of $\xi$ in \cite{CKRW}.
Finally, if $[CT(X,\delta), \alpha]$ has trivial Mackey
obstruction, then as explained in \cite[\S1]{PRW},
$CT(X,\delta)\rtimes_\alpha G$ has continuous trace and
has spectrum which is another principal torus bundle
over $Z$ (for the dual torus, $\widehat G$ divided
by the dual lattice).

Now consider the case when
\begin{equation}\label{h-1}
p_! \delta \ne 0 \in H^1(Z, \ZZ).
\end{equation}
It is still true as before that we can find an element
$[CT(X,\delta), \alpha]$ in $\Br_G(X)$ corresponding to
$\delta$. But there is no classical T-dual in this situation since
the Mackey obstruction \emph{can't} be trivial, because of
Theorem \ref{thm:n2facts}.  In fact, since any representative
$f\co Z\to \TT$ of a non-zero class in $ H^1(Z, \ZZ)$
must take on all values in $\TT$, there are necessarily
points $z\in Z$ for which the Mackey obstruction in
$H^2(\bbZ^2,\TT)\cong \TT$ is irrational, and hence the
crossed product $CT(X,\delta)\rtimes_\alpha G$ cannot
be type I. Nevertheless, we can view this crossed
product as a \emph{non-classical} T-dual to $(p,\delta)$.
The crossed product can be viewed as the algebra of
sections of a bundle of algebras (not locally trivial) over
$Z$, in the sense of \cite{DH}. The fiber of this
bundle over $z\in Z$ will be $C(p^{-1}(z), 
\cK(\cH)) \rtimes G \cong C(G/\ZZ^2, \cK(\cH)) 
\rtimes G \cong A_{f(z)} \otimes \cK(\cH)$,
which is Morita equivalent to the
twisted group {\Ca} $A_{f(z)} $ of the stabilizer group $\bbZ^2$ for
the Mackey obstruction class $f(z)$ at that point. In other words,
the  T-dual will be realized by a bundle of (stabilized)
\emph{noncommutative tori} fibered over $Z$. (See Figure 1.)

\begin{figure}[ht]
\includegraphics[height=1.5in]{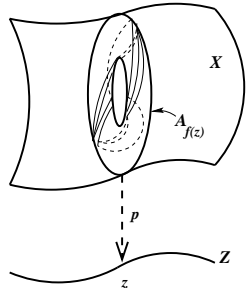}\\
\caption{In the diagram, the fiber over $z\in Z$ 
is the noncommutative torus $A_{f(z)}$,
which is represented by a foliated torus, with foliation 
angle equal to $f(z)$.}
\label{fig:folbundle}
\end{figure}

The bundle is not unique since there is no \emph{canonical}
representative $f$ for a given non-zero class in $H^1(X,\bbZ)$.
However, any two choices are homotopic, and the resulting
bundles will be in some sense homotopic to one another.
\end{proof}

As expected, our notion of T-duality comes with
isomorphisms in twisted $K$-theory and (cyclic) homology:
\begin{theorem}
\label{thm:Kiso}
In the situation of Theorem \textup{\ref{thm:main}}, if 
$X$ is a manifold, $H$ is an integral 3-form representing 
$\delta$ {\lp}in de Rham cohomology{\rp}, and we choose a smooth
model for $CT(X,\delta)$ {\lp}by taking a smooth bundle
over $X$ with fibers the smoothing operators{\rp}, we have
a commutative diagram
\begin{equation} \label{eq:Agn2}
\begin{CD}
K^\bullet(X, H)  @>T_!>\cong> K_{\bullet}(CT(X,\delta)\rtimes \RR^2)  \\
        @V{\Ch_H}VV          @VV{\Ch} V     \\
H^\bullet  (X, H)    @>T_*>\cong>  HP_{\bullet}
(CT(X,\delta)^\infty \rtimes \RR^2)
\end{CD}\end{equation}
where the horizontal arrows are isomorphisms,
$\Ch_H$ is the twisted Chern
character and $\Ch$ is the Connes-Chern character.

When $p_!\delta =0$ and there is a classical T-dual,
this reduces to a diagram like equation
\textup{\eqref{eq:TdualKthyn1}}, except that there is no
degree shift since the tori are even-dimensional.
\end{theorem}
\begin{proof}
This is done almost exactly as in \cite{BEMa}, so we will be brief.
We have the isomorphisms in $K$-theory
   $$
   \begin{array}{lcl}
K^{\bullet} (X, H )  &\cong &   K_\bullet(CT(X,\delta)) \\
&\cong & K_\bullet(CT(X,\delta)\rtimes \RR^2 ) \qquad
({\text{Connes-Thom isomorphism \cite{CThom}}}).
  \end{array}
   $$
   
We can also consider the smooth subalgebra  $CT(X,\delta)^\infty \rtimes G$.
The fiber at $z\in Z$ is given by
$C^\infty(p^{-1}(z), \cK^\infty(\cH)) \rtimes G 
\cong C^\infty(G/\ZZ^2, \cK^\infty(\cH)) \rtimes G \cong
A_{f(z)}^\infty \otimes \cK^\infty(\cH),$
where $\cK^\infty(\cH)$ is the algebra of smoothing operators on $\cH$
and $A_{f(z)}^\infty$ is the smooth noncommutative torus with 
multiplier
equal to $f(z)$.

Then we have the isomorphisms
$$
   \begin{array}{lcl}
    H^{\bullet} (X, H) &\cong & HP_\bullet(CT(X,\delta)^\infty )  \\
  &\cong &  HP_\bullet(CT(X,\delta)^\infty\rtimes \RR^2 )
\qquad ({\text{ENN-Thom isomorphism \cite{ENN}}}).
    \end{array}
$$

It is well known that the Chern character is compatible with the 
isomorphisms in $K$-theory and cohomology, from which 
the commutativity of the diagram in \eqref{eq:Agn2}
follows.
\end{proof}

\begin{remark}
\label{rem:remdualH}
The reader might wonder what happened to the dual H-flux $H^\#$
in the context of Theorem \ref{thm:main}(2). It doesn't really
make sense as a cohomology class or differential form since
the nonclassical T-dual is not a space; rather, it is subsumed
in the noncommutative structure of the dual.

Now let us describe the relationship between our Definition
\ref{def:classicaldual} and Theorem \ref{thm:main} and the
corresponding notions in \cite{BHMb}.  If the pair $(p\co X
\to Z, \delta)$ is T-dualizable in the sense of \cite{BHMb},
that means $\delta$ is represented by a closed 3-form $H$,
such that $\iota_\Xi H = p^* \widehat F(\Xi)$, for some
integral closed 2-form $\widehat F$ with values in the dual of
$\Fg$, the Lie algebra of $T$, and for all
$\Xi\in\Fg$. This essentially means
that when we integrate $H$ over the fibers of $p_1$,
where $X \xrightarrow{p_1} Y \xrightarrow{p_1} Z$ is a factorization
of $p$ into two circle bundles, then the resulting 2-form
is pulled back from $Z$. This implies in turn that
integrating $H$ over the fibers of $p$ gives $0$, 
which is the condition $p_![H]=0$. (We do not need to
worry about torsion in cohomology since $p_!\delta$ lies in
$H^1(Z,\bbZ)$, which is always torsion-free.) Thus
the condition in our Theorem \ref{thm:main}(1) is satisfied.

Conversely, suppose our condition $p_!\delta=0$ is satisfied,
so we have a classical T-dual $(p^\#\co X^\# \to Z, \delta^\#)$.
The condition of \cite{BHMb} that $\iota_\Xi H = p^* \widehat F(\Xi)$,
for some  closed integral 2-form $\widehat F$ with values in the dual of
$\Fg$ and for all
$\Xi\in\Fg$, will follow from the fact that since $p_!\delta=0$ (and
we can divide out by trivial cases where $\delta$ is pulled
back from $Z$), $\delta$ comes from the $E^{2,1}_\infty$
subquotient of $H^3(X,\bbZ)$.
\end{remark}

\section{Examples: torus bundles and noncommutative torus 
bundles over the circle}
\label{sec:ex}

A famous example of a principal torus bundle with non T-dualizable 
H-flux
is provided by $\TT^3$, considered as the trivial $\TT^2$-bundle over 
$\TT$,
with $H$ given by $k$ times the volume form on $\TT^3$.  $H$ is non 
T-dualizable
in the classical sense since $p_![H]\ne 0$. Alternatively,
there are no non-trivial  $\TT^2$-bundles 
over $\TT$, since
$H^1(\TT,\underline{\TT^2}) \cong H^2(\TT,\ZZ^2) =0$, that is, there
is no way to dualize the H-flux
by a (principal) torus bundle over $\TT$.

This example is covered by Theorem \ref{thm:main}(2) and
by Theorem \ref{thm:Kiso}.
The T-dual is realized by a bundle of stabilized
{\em noncommutative tori} fibered over $\TT$.
In fact the construction of the non-classical T-dual
in this case is a special case of the construction in the
proof of Proposition \ref{prop:Mobstrfortrivbundle}, but
we repeat the details since we can make things more explicit.
Let $\cH = L^2(\TT)$ and consider the the projective unitary 
representation
$\rho_\theta : \ZZ^2 \to \PU(\cH)$ given by the first $\ZZ$ factor 
acting by
multiplication by $z^k$ (where $\TT$ is thought of as the unit circle 
in $\CC$)
and the second $\ZZ$ factor acting by translation by $\theta \in \TT$.
Then the Mackey obstruction of $\rho_\theta$ is $\theta \in \TT \cong 
H^2(\ZZ^2, \TT)$.
Let $\ZZ^2$ act on $C(\TT, \cK(\cH))$ by $\alpha$, which is given at 
the  point $\theta$ by $\rho_\theta$. Define the $C^*$-algebra,
  $$
  \begin{array}{lcl}
  B &=& {\Ind}_{\ZZ^2}^{\RR^2}\left(C(\TT, \cK(\cH)), \alpha\right) 
\\[+7pt]
  &=& \left\{ f: \RR^2 \to C(\TT, \cK(\cH)):
  f(t+g) = \alpha(g)( f(t)),\;\; t\in \RR^2, g\in\ZZ^2 \right\}.
  \end{array}
  $$
That is, $B$ is a mapping torus of a $\ZZ^2$-action on $C(\TT, \cK(\cH))$.
Then $B$ is a continuous-trace {\Ca} having spectrum $\TT^3$,
having an action of $\RR^2$ whose induced action on the spectrum of $B$
is the trivial bundle $\TT^3\to \TT$.  The crossed product algebra
$B\rtimes \RR^2 \cong C(\TT, \cK(\cH))\rtimes \ZZ^2$ has fiber
over $\theta \in \TT$ given by
$\cK(\cH) \rtimes_{\rho_\theta} \ZZ^2 \cong 
A_\theta \otimes \cK(\cH),$
where $A_\theta$ is the noncommutative $2$-torus.
In fact, the crossed product  $B\rtimes \RR^2$
is Morita equivalent to $C(\TT, \cK(\cH))\rtimes\bbZ^2$
and is even isomorphic to the stabilization of this
algebra (by \cite{Green}).  Thus $B\rtimes \RR^2$ is isomorphic
to $C^*(H_{\bbZ} ) \otimes \cK$, where $H_{\bbZ}$ is the
integer Heisenberg-type group,
\[
H_{\mathbb Z} = \left\{   \begin{pmatrix} 1 & x & \frac{1}{k}z \\
0 & 1 & y\\
0 & 0 &       1
\end{pmatrix} : x, y, z \in \mathbb Z \right\},
\]
a lattice in the usual Heisenberg group $H_{\bbR}$ (consisting
of matrices of the same form, but with $x,\,y,\,z\in\bbR$).
Then we have the isomorphisms in $K$-theory
\[
  \begin{array}{lcl}
  K_\bullet(B) & = & K^\bullet(\TT^3,k\, d\vol)
 \qquad (\text{definition})\\
  &\cong & K_\bullet(B\rtimes \RR^2 ) \qquad 
({\text{Connes-Thom isomorphism}})\\
  &\cong & K_\bullet(C^*(H_{\bbZ} )) \qquad ({\text{above 
identification}})\\
  &\cong & K_\bullet(H_{\bbR}/H_{\bbZ}) \qquad ({\text{Baum-Connes 
conjecture}})\\
   &\cong & K^{\bullet +1} (H_{\bbR}/H_{\bbZ}) 
\qquad ({\text{Poincar\'e duality}}).
   \end{array}
\]
where we observe that the Heisenberg nilmanifold $H_{\bbR}/H_{\bbZ}$
(which happens to be the classifying space $BH_{\mathbb Z} $) is a 
circle bundle
over $\TT^2$ with first Chern class equal to $k dx \wedge dy$.

Notice that as far as $K$-theory is concerned, the T-dual
of $(T^3,k\,d\vol)$ can also be taken to be the
nilmanifold $H_{\bbR}/H_{\bbZ}$ with the trivial $H$-field.
This is a \emph{non-principal} $T^2$-bundle over $S^1$.
But a better model for a non-classical T-dual
is simply the group {\Ca} of $H_{\bbZ}$.

We can also consider the smooth subalgebra 
$B^\infty$ of $B$ defined by
  $$
  \begin{array}{lcl}
  B^\infty &=& {\rm Ind}_{\ZZ}^{\RR}\left(C^\infty(\TT, 
\cK^\infty(\cH)), \alpha\right) \\[+7pt]
  &=& \left\{ f: \RR^2 \to C^\infty(\TT, \cK^\infty(\cH)):
  f(t+g) = \alpha(g)( f(t)),\;\; t\in \RR^2, g\in\ZZ^2 \right\}.
  \end{array}
  $$
where $\cK^\infty(\cH))$ denotes the algebra of smoothing operators on 
$\TT$.
$B^\infty\rtimes \RR^2 \cong C^\infty(\TT, \cK^\infty(\cH))\rtimes 
\ZZ^2$ has fiber
over $\theta \in \TT$ given by
$\cK^\infty(\cH) \rtimes_{\rho_\theta} \ZZ^2 \cong A_\theta^\infty 
\otimes \cK^\infty(\cH),$
where $A_\theta^\infty$ is the smooth noncommutative torus and the 
tensor product
is the projective tensor product. In this case, the
crossed product $B^\infty\rtimes \RR^2 \cong \cS(H_{\mathbb Z} )
\otimes \cK^\infty(\cH)$, where $\cS(H_{\mathbb Z} )$ is the rapid 
decrease algebra.
Then we have the isomorphisms
$$
  \begin{array}{lcl} 
HP_\bullet(B^\infty ) & = & H^{\bullet} (\TT^3, k\, d\vol ) \qquad 
({\text{definition}})\\
&\cong & HP_\bullet(B^\infty\rtimes \RR^2 ) 
\qquad ({\text{ENN-Thom isomorphism}})\\
  &\cong & HP_\bullet(\cS(H_{\mathbb Z} )) \qquad ({\text{Above 
identification}})\\
  &\cong & H_\bullet(H_{\bbR}/H_{\bbZ}) \qquad ({\text{Cyclic homology 
Baum-Connes conjecture}})\\
   &\cong & H^{\bullet +1} (H_{\bbR}/H_{\bbZ}) \qquad ({\text{Poincar\'e 
duality}})\\
      \end{array}
  $$
where $HP_\bullet$ denotes the periodic cyclic homology, which
  is stable under (projective) tensor product with $ \cK^\infty(\cH)$
  and $H_\bullet$,  $H^{\bullet} $ denote the $\bbZ_2$-graded
  homology and cohomology respectively.

Finally, T-duality can be expressed  in this case by the following 
commutative diagram,
\begin{equation} \label{eq:Ag}
\begin{CD}
K^\bullet(\TT^3, k\, d\vol)  @>T_!>> K_{\bullet}(C^*(H_\bbZ))  \\
       @V{\Ch_H}VV          @VV{\Ch} V     \\
H^\bullet  (\TT^3, k\, d\vol)    @>T_*>>  HP_{\bullet} 
(H^\infty(H_{\mathbb Z} ))
\end{CD}\end{equation}
where $H=k \,d\vol$, $\Ch_H$ is the twisted Chern 
character
and $\Ch$ is the Connes-Chern character.

\section{Concluding remarks}
\label{sec:concl}

In this paper, we have only dealt with complex {\Ca}s
and  complex $K$-theory, which are relevant for type II
string theory.  In principle, most of what we have
done should also extend to the type I case, which
involves real $K$-theory. However, one has to be
careful. Since $T$-duality is related to the Fourier
transform, and since the Fourier transform of a real
function is not necessarily real, a theory of T-duality
in type I string theory necessarily involves $KR$-theory, or Real
$K$-theory in the sense of Atiyah \cite{Akr}.  The correct
notion of twisted $KR$-theory is that of $K$-theory
of real {\CTalg}s in the sense of \cite[\S3]{RosCT}.
What complicates things is that such algebras
are built out of {\CTalg}s of real, quaternionic,
and complex type (locally isomorphic to
$C(X,\cK_\bbR)$, $C(X,\cK_\bbH)$, and $C(X,\cK_\bbC)$, 
respectively). Even if one's original interest is
in algebras of real type, passage to the T-dual
will often involve algebras of the other types.

One possibility suggested by the example in section
\ref{sec:ex} is that there is a good theory of
T-duality for arbitrary torus bundles with H-fluxes,
that doesn't require going to a category of noncommutative 
bundles, but that it is necessary to include the possibility
of non-principal bundles. We have seen that there is
a sense in which the Heisenberg nilmanifold (with trivial
$H$-field) can be viewed as a T-dual to $T^3$ with
a non-trivial $H$-field. (This is literally true
in the sense of \cite{BEMa} if we think of both manifolds
as $\TT$-bundles over $T^2$, rather than as $T^2$-bundles
over $S^1$.)

It is of course a little disappointing that our main
theorem only applies when the fibers of the torus
bundle are $2$-dimensional. From Theorem \ref{thm:CKRWimproved},
it is not even clear if the map $\Br_G(X)\to H^3(X,\bbZ)$
is surjective when $n=\dim G >2$. However, the methods of
this paper should apply on the image of this map.

\end{document}